# Generalized Epsilon-Near-Zero Polaritons on Uniaxial Metasurfaces

Francisco Javier Alfaro-Mozaz,[1,*] and Iñigo Liberal[1,†]

[1] *Department of Electrical, Electronic and Communications Engineering, Institute of Smart Cities (ISC), Public University of Navarre (UPNA), 31006 Pamplona, Spain*

Epsilon-near-zero (ENZ) thin films facilitate strong light-matter interactions with a widespread impact in nonlinear, quantum and thermal photonics. Here, we extend the scope of thin film ENZ modes by elucidating the generalized polaritonic modes emerging from anisotropy and near-zero permittivity. Through a theoretical investigation of generalized ENZ polaritons on silicon carbide (SiC) metasurfaces, we reveal a complex polaritonic landscape, consisting of a seven-region phase diagram. We show that the complex interplay between material properties and geometry changes the nature of the modes, and we clarify when ENZ modes with closed and open isofrequency curves (IFCs) can be observed, and how they coexist with surface phonon-polaritons and hyperbolic modes. The associated photonic topological transitions are accompanied by phase velocity sign reversals and induce dramatic modifications of near-field profiles and Purcell enhancement factors. Our work merges insights from anisotropic and ENZ effects, enabling enhanced control over polariton behavior with broad implications for subwavelength optics, material design, nonlinear effects, thermal emission, and quantum technologies.

## I. INTRODUCTION

Epsilon-near-zero (ENZ) thin films –deeply subwavelength layers with a near-zero permittivity [1]– have recently had a profound impact on some of the most prominent areas of nanophotonics. First, they exhibit strong and ultrafast nonlinearities [2,3] entering into the nonperturbative regime [4], which has important implications for the broad field of nonlinear optics [5,6]. For example, they have powered nonperturbative and ultrafast Kerr effects [2,3], frequency conversion [7], strong second [8], third [9], and high-harmonic generation [10], all-optical switching [11], modulation [12] and the generation of polarization-entangled Bell states [13]. For similar reasons, ENZ thin films are the major driving force behind spatiotemporal metamaterials [14–16], which has enabled demonstrations of time-refraction [17,18], double time-slits experiments [19], superluminal synthetic motion [20], ultrafast mirrors [21], and spatiotemporal knife detectors [22]. In addition, the optical response of ENZ thin films based on doped semiconductors can be electrically tuned at the dielectric to metal transition, leading to fast reconfigurable metasurfaces [23] and tunable heat transfer [24]. Moreover, magnetized ENZ thin films have facilitated the first experimental reports on the violation of Kirchhoff's law of thermal radiation [25,26].

In addition, ENZ thin films support electromagnetic modes pinned to the ENZ wavelength, referred to as "ENZ modes" [27–29]. These modes are characterized by large wavevectors, high field confinement within the film and strong dispersion. These modes provide strong coupling to other electromagnetic excitations [30–35] and thin films supporting ENZ and plasmon polaritons maximize Casimir-induced accelerations [36].

The distinct electromagnetic behavior of ENZ photonics stems from different factors including extreme boundary conditions [37], high intrinsic impedance [38,39], and near-zero group velocity [7], with its ultimate performance limited by material loss [6] and surface roughness [40].

In parallel with the development of the field of ENZ photonics, there is a growing interest in polaritons supported by anisotropic thin films. Such films, - whether naturally anisotropic (*e.g.*, h-BN, MoO$_3$, *β*-Ga$_2$O$_3$ [41–45]) or artificially structured into subwavelength metasurfaces [46–50], exhibit rich polaritonic phenomena such as hyperbolic, ghost and shear polaritons. In particular, it has been demonstrated that an h-BN metasurface can exhibit a hyperbolic in-plane response by laterally nanostructuring it in an array of subwavelength ribbons [49].

Here, we bridge aspects of ENZ photonics and anisotropic polaritons by studying ENZ polaritons in silicon carbide (SiC) uniaxial metasurfaces, and the photonic topological transitions [51] associated with generalized ENZ regimes. It is found that the nature of the polaritonic modes qualitatively changes with the induced anisotropy, and we identify in which phases ENZ modes still exist. Such complexity emerges from the interplay between material (*e.g.*, ENZ points with negative to positive transitions) and structural (ENZ modes, surface phonon polaritons (SPhPs) and their hybridization) considerations. Specifically, we map the complete polaritonic phase diagram of the homogenized metasurface (see Fig. 1d), realized using a strategy similar to that of Refs. [49,50].

*javier.alfaro@unavarra.es

†inigo.liberal@unavarra.es

Our classification framework, which explicitly connects tensorial permittivity components to distinct polaritonic phases, recasts specific polaritonic modes as ENZ polaritons bringing to light their associated topological transitions. By linking permittivity to polaritonic phases, our findings provide specific guidelines for engineering metasurfaces with tailored ENZ electromagnetic responses, enabling precise control of polaritonic propagation and local density of states.

## II. MATERIAL PROPERTIES AND REGIMES

### A. SiC thin films

Before delving into anisotropic metasurfaces, we revisit the theory of confined modes in SiC thin films, highlighting two often overlooked properties. To this end, we consider a 100 nm thick SiC layer in a vacuum environment ($\epsilon = 1$, see Fig. 1a). SiC is chosen since it sustains long-lived phonon-polaritons in its Reststrahlen band in the mid-IR [52–54], as well as for the availability of high-quality thin films, which makes it a suitable platform for investigating anisotropic polaritonic phenomena. The film thickness of 100 nm is sufficiently small to support an ENZ polariton near its LO phonon frequency [27]. Finally, the use of a vacuum environment simplifies the model and is consistent with experiments where long-lived phonon-polaritons were visualized [53,55].

The SiC permittivity is given by:

$$\epsilon_{SiC}(\omega) = \epsilon_{\infty}\left(\frac{\omega^2 - \omega_p^2 + i\gamma\omega}{\omega^2 - \omega_0^2 + i\gamma\omega}\right), \quad (1)$$

with the parameters [40,56] indicated in Table I. The SiC Reststrahlen band covers the mid-IR region between 10.3 and 12.55 μm (background colored region in Fig. 1b). The dispersion of the modes is given by [57]:

*javier.alfaro@unavarra.es
†inigo.liberal@unavarra.es

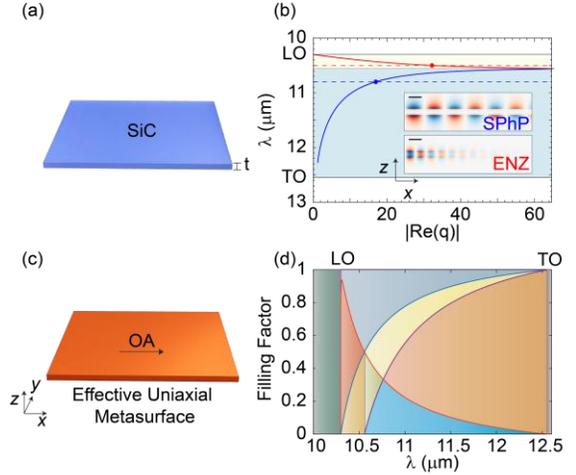

Figure 1. **Uniaxial metasurface description**. (a) t = 100 nm thick SiC slab (b) Dispersion relations for the SPhP (blue solid line) and ENZ polaritons (red solid line). Background color indicates the ENZ (light yellow) and SPhP (light blue) regions. Inset: Vertical component of the electric field of the modes propagating from left to right at the wavelengths marked by dashed lines. Bar length: 200 nm. (c) Homogenized uniaxial metasurface. (d) Polariton region diagram as a function of wavelength and filling factor (Eq. 3), background colors indicate different regions. OA: Optical Axis.

$$\begin{aligned}
\text{ENZ} &\coloneqq \tanh\left(\frac{\kappa_2 t}{2}\right) = -\epsilon_{SiC}\frac{\kappa_1}{\kappa_2}, \\
\text{SPhP} &\coloneqq \coth\left(\frac{\kappa_2 t}{2}\right) = -\epsilon_{SiC}\frac{\kappa_1}{\kappa_2},
\end{aligned} \quad (2)$$

where $\kappa_1 = \sqrt{q^2 - k_0^2}$ and $\kappa_2 = \sqrt{q^2 - \epsilon_{SiC}k_0^2}$ are the $z$-component of the wavevectors, outside and inside the layer respectively, and $q$ the normalized in-plane wavevector. The SPhP dispersion curve (also called $\omega^-$ [57]) starts at $\omega_0$ and its wavevector increases with frequency up to its asymptote at $\epsilon_{SiC}(\omega) = -1$. The ENZ polariton, (also called $\omega^+$ [57,58]) has a flattened dispersion profile contained in the frequency range fulfilling $-1 < \epsilon_{SiC} < 0$ (Fig. 1d). Both polaritons are in-plane isotropic (same wavelength in each in-plane direction) TM-polarized waves.

The first often overlooked property is that the phase velocity of the ENZ polariton is negative. See Supp. Mat. for analytical and numerical calculations supporting this fact, which was noted in early works on plasmonics [58,59]. On the other hand, the phase velocity is positive for the SPhP, leading to an interesting interplay in their potential hybridization.

TABLE I. SiC Drude-Lorentz parameters

| $\epsilon_\infty$ | $\omega_p [\mu m]$ | $\omega_0 [\mu m]$ | $\gamma [cm^{-1}]$ |
|---|---|---|---|
| 5.726 | 10.3 | 12.55 | 1.753 |

The group velocity is positive for the two modes. The SPhP electric field (top inset of Fig. 1b) is mostly located in air, decaying towards free space from the air/SiC boundaries. In contrast, the ENZ polariton electric field (bottom inset of Fig. 1b) is largely located inside the SiC layer and is dominated by its vertical electric field.

The second overlooked property is that, despite the fact that the ENZ polariton emerges in the ultrathin film limit [27,30], its response cannot be recovered in the infinitesimally-thin or sheet limit (see Supp. Mat.). Therefore, even operating in an ultra-thin limit the properties of ENZ modes are fundamentally different from those of 2D photonics.

## B. Metasurface based on SiC

To obtain an anisotropic optical response, we pattern the SiC flake into ribbons of width $w$ and gap $g$, for a total period $L = w + g$ (Fig. 2a). In order to provide a first-order theoretical description of the patterned metasurface, we estimate the effective permittivities using a simple homogenization model [49,60,61]:

$$\epsilon_x = \left[\frac{(1-f)}{\epsilon_{SiC}} + f\right]^{-1}, \quad (3)$$
$$\epsilon_y = \epsilon_z = \epsilon_{SiC}(1-f) + f,$$

where $f = g/L$ is the filling factor (Fig. 2a). Thus, the effective permittivity corresponds to that of an in-plane uniaxial medium with the optical axis in the x-direction and components $\hat{\epsilon} = \text{diag}(\epsilon_x, \epsilon_y, \epsilon_y)$. We stress that while this approach is based on standard effective medium approximations –which do not fully capture nonlocal effects and the precise polaritonic mode distribution in such subwavelength structures– it nonetheless provides a useful theoretical framework for exploring the anisotropic optical response of arbitrary uniaxial metasurfaces. The existence of the predicted optical modes is confirmed with numerical simulations of the ribbon system (see Discussion and Supp. Mat.).

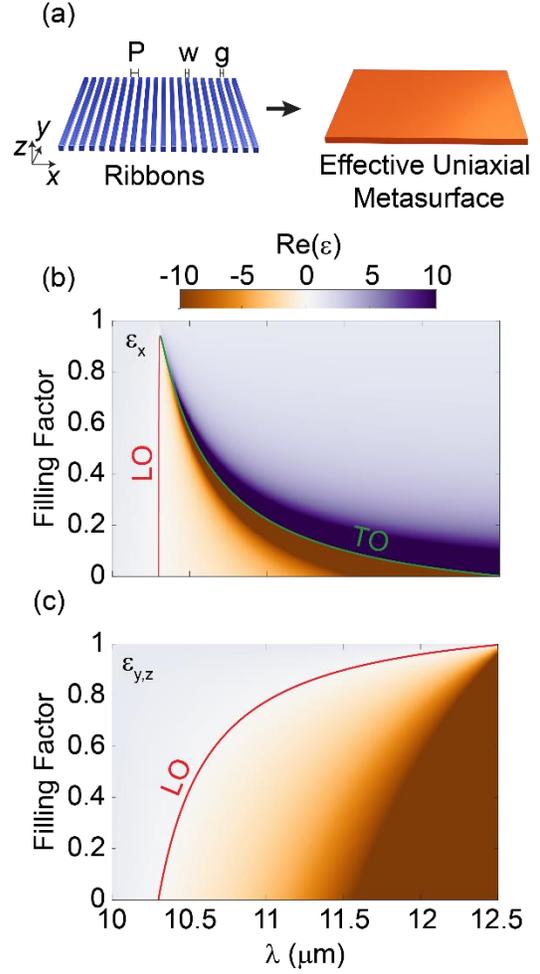

Figure 2. **Real part of the permittivity components as a function of filling factor and wavelength**. (a) schematic of the homogenization of a ribbon array (b) $\epsilon_x$, (c) $\epsilon_{y,z}$. Red solid line indicates the effective LO phonon and green solid line the effective TO phonon. Colors have been saturated for better visibility.

The values of the components of the permittivity tensor, for all filling factors across the Reststrahlen band of SiC are plotted in Fig. 2. In pristine SiC, the real part of the permittivity becomes negative at the LO phonon at 10.3 μm, and a rapid variation and a change of sign of it occurs at the TO phonon wavelength at 12.55 μm. On the metasurface, the TO phonon wavelength of $\epsilon_x$ component of the metasurface (green line in Fig. 2b) lowers for increased filling factors until it collapses to the LO phonon wavelength (red lines in Fig. 2b) as the filling factor reaches one. In contrast, the TO phonon wavelength is constant for the $\epsilon_y$ component. The $\epsilon_y$ effective LO phonon wavelength (red line in Fig. 2c) however, increases until it collapses to the TO phonon wavelength at filling factors tending to one. The real part of the two permittivity components traditionally divides the polaritonic phase diagram into four distinct zones: (i) $\epsilon_x, \epsilon_y < 0$, (ii) $\epsilon_x, \epsilon_y > 0$, (iii) $\epsilon_x > 0, \epsilon_y < 0$, (iv) $\epsilon_x < 0, \epsilon_y > 0$. However, as anticipated, additional polaritonic phases are identified in the generalized ENZ regimes. In particular, the complete polaritonic phase diagram consists of three more regions as new boundaries appear at $\epsilon_y = -1$ and $\epsilon_x \epsilon_y = 1 \cup \epsilon_{x,y} < 0$ for a total of seven distinct regions inside the Reststrahlen band of SiC. The outline of the polaritonic phase diagram is shown in Fig. 1d and the precise boundary values are presented in Table II. Next, we analyze the existence of ENZ modes in these different regions.

*javier.alfaro@unavarra.es

†inigo.liberal@unavarra.es

TABLE II. Regions of the polaritonic phase diagram of the metasurface. All permittivity components refer to their real parts.

| Region | Conditions |
|---|---|
| I | $\epsilon_x < 0, \epsilon_y > 0, \epsilon_x \epsilon_y < 1$ |
| II | $\epsilon_x < 0, -1 < \epsilon_y < 0, |\epsilon_x \epsilon_y| < 1$ |
| III | $\epsilon_x, \epsilon_y > 0$ |
| IV | $\epsilon_x < 0, -1 < \epsilon_y < 0, |\epsilon_x \epsilon_y| > 1$ |
| V | $\epsilon_x > 0, -1 < \epsilon_y < 0$ |
| VI | $\epsilon_x > 0, \epsilon_y < -1$ |
| VII | $\epsilon_x < 0, \epsilon_y < -1, \epsilon_x \epsilon_y > 1$ |

### III. POLARITONIC MODES IN EACH REGION

The dispersion of the polaritonic modes in the uniaxial metasurface, surrounded by a symmetric environment of permittivity $\epsilon_s = 1$, in the high-momentum approximation (see Supp. Mat.), is given by [62,63]:

$$q(\omega, \alpha) = \frac{\rho^\pm}{k_0 t}[2\operatorname{atan}(\xi) + \pi l], \quad (4)$$

where $q(\omega, \alpha)$ is the normalized in-plane wavevector $q(\omega, \alpha) = \frac{k_\parallel(\omega, \alpha)}{k_0(\omega)}$ at a direction $\alpha$ with respect to the optical axis (x-axis). $t$ is the thickness of the layer, $\rho^\pm = \pm i \sqrt{\frac{\epsilon_y}{\epsilon_x \cos^2 \alpha + \epsilon_y \sin^2 \alpha}}$, $\xi = \frac{\rho^\pm}{\epsilon_y}$, and $l \in \mathbb{N}_0$. On uniaxial media, the polaritonic modes are purely TM polarized [60]. We assume that $\operatorname{Re}(\boldsymbol{q}) \parallel \operatorname{Im}(\boldsymbol{q})$, which is a good approximation for the purposes of the article [63]. The physical solutions of the propagating modes correspond to $\operatorname{Re}(\boldsymbol{q}) > 0$, $|\operatorname{Re}(\boldsymbol{q})| > |\operatorname{Im}(\boldsymbol{q})|$ (see Supp. Mat.). Moreover, physical solutions are purely real, $\boldsymbol{q} \in \mathbb{R}$, if $\epsilon_{x,y} \in \mathbb{R}$ (see Supp. Mat.). In the following section we will analyze the polaritonic modes of the metasurface in the lossless limit using the previous property, and compare it with the dispersion relation, Eq. (4), and Transfer Matrix Calculations [64] in the case with losses. To calculate the phase velocity, we will assume small losses in $\hat{\epsilon}$. Solutions with $\operatorname{Im}(\boldsymbol{q}) > (<)0$ correspond to modes with positive (negative) phase velocity $v_p$.

By analyzing Eq. (4) (see Supp. Mat.) we describe and interpret *all* polaritonic modes in each of the regions (Table II). This systematic treatment serves three main goals: (i) to reveal not explicitly addressed properties of known modes and regions, such as phase velocity and field confinement, (ii) to characterize polaritonic regimes that had not been identified within the ENZ framework, (iii) to place all these modes within a unified polaritonic phase diagram. In doing so, we derive each region's boundaries, the openness and closedness of its IFCs, the asymptote angles of the IFC, the modes phase velocity and electric field distributions. Numerical simulations [65] are performed to corroborate the results and to illustrate the real-space field distributions induced by point sources.

#### 1. Region I, ENZ region with open IFC polaritons

This region is contained between the LO phonon frequency and the $\epsilon_{SiC} = -1$ frequency (Fig. 3g), where the permittivity tensor is hyperbolic, that is, sign Re $(\epsilon_x) \neq$ sign Re $(\epsilon_y)$. Interestingly, due to the vanishing value of the permittivity, even a small perturbation (filling factor) on the original system can lead to a hyperbolic permittivity tensor near the LO phonon frequency of SiC. The physical solutions of Eq. (4) correspond to $\rho^+(\alpha) \in \mathbb{R}^+$, which traces a hyperbola. These solutions represent a family of modes of index $l \in \mathbb{N}_0$ characterized by open IFCs with its vertex located on the x-axis, having higher momenta the higher the mode index $l$. The asymptote of these curves occurs at the branch cut of $\rho^+$, i.e. when $\rho^+ \to \infty$:

*javier.alfaro@unavarra.es
†inigo.liberal@unavarra.es

$$\theta_1 = \operatorname{atan}\sqrt{-\frac{\epsilon_x}{\epsilon_y}}, \quad (5)$$

which leads to $q(\alpha_a, \omega_0) \to \infty$. The phase velocity of these modes is positive. The field profile of these modes displays an increasing number of nodes in $E_z$ such that $N_{nodes}(E_z) = l + 1$ (similar to that of Fig. 4c). A complete set of field profiles can be found in the Supp. Mat.). The electric fields induced by a point source over the metasurface (Fig. 3d) show polaritons with concave wavefronts propagating along the *x*-axis, illustrating the polaritonic propagation induced by the open IFCs. Crucially, while the permittivity components are near-zero, $|\operatorname{Re}(\epsilon_x)\operatorname{Re}(\epsilon_y)| < 1$, the cross-sectional field distribution (Fig. 4c) and phase velocity of the modes here are not characteristic of conventional ENZ modes. We conclude that Region I does not support ENZ modes; instead, the observed modes are in-plane hyperbolic volume modes (HVMs) [41,63]. This region exemplifies the interplay between materials and geometry, which does not simply induce an anisotropy into the supporting modes but change its basic properties, in this case, forbidding the existence of ENZ modes.

## *2. Region II, negative permittivity region with anisotropic ENZ polaritons*

This region corresponds to that of an anisotropic ENZ material, with both components of the permittivity tensor negative $\operatorname{Re}(\epsilon_x) < 0, -1 < \operatorname{Re}(\epsilon_y) < 0$, $|\operatorname{Re}(\epsilon_x)\operatorname{Re}(\epsilon_y)| < 1$. The physical solutions of Eq. (4) correspond to $\rho^- \in \mathbb{I}^-$, with $\operatorname{Im}(\rho^-)$ tracing an ellipse. Since $\operatorname{atan}(\xi) \in \mathbb{I}^+ > 1i$, we can rewrite it as $i \cdot \operatorname{atanh}(-i\xi) - \pi$. A crucial result that follows is that only the $l = 1$ mode exists, which yields a closed IFC (see Fig. 3b) and negative phase velocity. Fig. 3e illustrates the polaritonic mode propagating across all angles with different wavelengths at each angle, confirming a single closed IFC anisotropic polaritonic mode. The mode's cross-sectional electric field is mainly vertical and confined to the metasurface layer (similar to Fig. 4a). Moreover, this region also covers the case of a pristine SiC layer at the ENZ region. Since in this region a single, negative phase velocity mode exists, the metasurface acts as an anisotropic negative-index layer [59]. Hence, Region II supports an anisotropic, closed IFC ENZ mode, generalizing the conventional isotropic ENZ polariton [27] to anisotropic metasurfaces.

## *3. Region III, anisotropic dielectric region*

This region corresponds to an anisotropic dielectric layer, with $\operatorname{Re}(\epsilon_{x,y}) > 0$. In a symmetric environment, thin film dielectrics can support waveguide modes with a wavevector similar to that of free-space light [66,67]. Consequently, it cannot guide high-*q* modes, which are the scope of the article.

*javier.alfaro@unavarra.es

†inigo.liberal@unavarra.es

## *4. Region IV, negative permittivity region with open IFC polaritons*

This region corresponds to an all-negative permittivity tensor $\operatorname{Re}(\epsilon_x) < 0$, $-1 < \operatorname{Re}(\epsilon_y) < 0$, $|\operatorname{Re}(\epsilon_x)\operatorname{Re}(\epsilon_y)| > 1$, with the $\operatorname{Re}(\epsilon_y)$ component being epsilon-near-zero. In this case, two distinct modes, corresponding to two sets of physical solutions appear. The first corresponds to $\rho^- \in \mathbb{I}^+$ with $\operatorname{atan}(\xi) \in \mathbb{I}^+ < 1i$ (See Supp. Mat.). The solution corresponds to $l = 0$ mode, which has an open IFC with its vertex at $\alpha = 0$. The IFC asymptote occurs at the branch cut of the arctangent term at $\xi = i$, at an angle:

$$\theta_2 = \arctan\sqrt{\frac{\epsilon_x\epsilon_y - 1}{1 - \epsilon_y^2}}. \quad (6)$$

Surprisingly, this asymptote is not derived from the *effective refractive index* [60] (proportional to $\rho^2$) of the metasurface, since $\rho^-$ describes an ellipse, but from the branch cut of the arctan term, which appear due to the planar geometry of the system. This mode has a positive phase velocity and its electric field distribution is that of a SPhP (similar to Fig. 4b) in a SiC layer

The second mode corresponds to $\rho^+ \in \mathbb{I}^-$ with the $\xi \in \mathbb{I}^+ > 1i$ (see Supp. Mat.). The solution corresponds to the $l = 1$ mode, which is an open curve with its vertex at $\alpha = \frac{\pi}{2}$. Its asymptote occurs where the argument of the atan equal to one, thus corresponding to Eq. (6). The phase velocity of the mode is negative, and its cross-sectional electric field distribution (similar to Fig. 4a) resembles that of the ENZ mode on the SiC layer. Therefore, this region supports open IFCs ENZ modes.

Hence, in this region two modes coexist at complementary angles (Fig. 3c). Its isofrequency curves (IFCs) are open, and its phase velocity direction is opposite, with electric field distributions corresponding to that of an ENZ polariton and that of an SPhP. This is confirmed by the electric fields induced by a dipole over the metasurface (Fig. 3f), which shows two different concave-wavefront waves propagating at perpendicular directions, while sharing the same asymptotic angle. We note that a similar dispersion was found in [62,68], here we have expanded the discussion and analyzed it thoroughly from the perspective of ENZ photonics. Surprisingly, these IFCs occur in a regime where both permittivity components are negative, hence demonstrating that $\hat{\epsilon}$-negative metasurfaces can sustain open IFC polaritons in the ENZ regime.

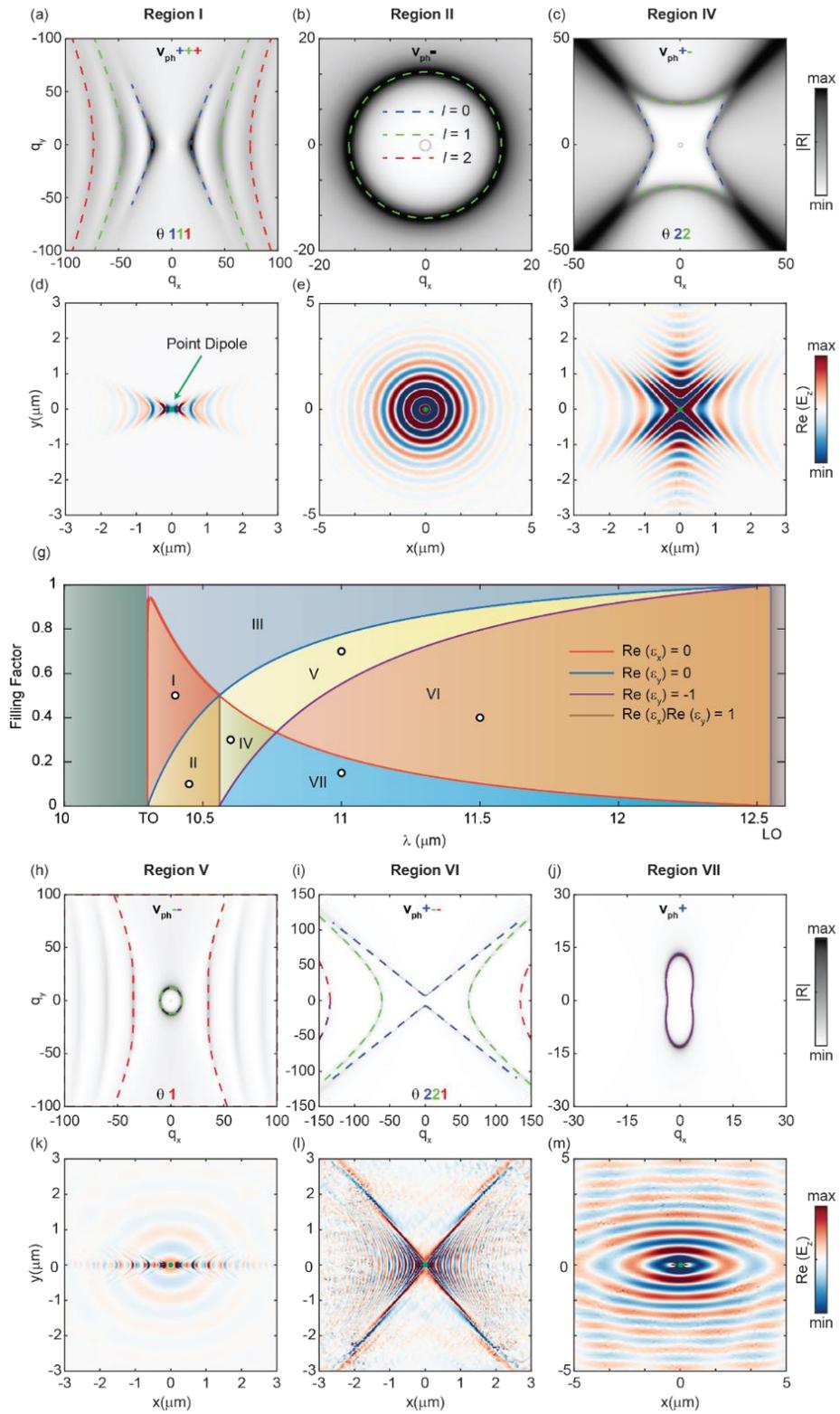

*javier.alfaro@unavarra.es
†inigo.liberal@unavarra.es

Figure 3. **Polaritonic modes and polaritonic phase diagram**. (a-c,h-j) Calculated isofrequency curves in momentum space for regions I, II, IV, V, VI, and VII. Dashed lines indicate the solutions of Eq. (4), while the color map shows the calculated reflectivity [59]. The symbols '+' and '-' following $v_p$ indicate the phase velocity sign of each mode, with blue, green, and red corresponding to $l = 0, 1$, and 2 respectively. For open IFC modes, the asymptotic angles are labeled as "1" and "2", corresponding to $\theta_1$ and $\theta_2$, given by Eqs. (5,6) of the main text. (d-f, k-m) Full-wave simulations of the real-space electric field distribution, Re($E_z$), over the metasurface induced by a point dipole source (green dot) at the origin. (g) Outline of the polaritonic phase diagram, showing its regions as a function of filling factor and wavelength, with points marking the representative λ and filling factors values analyzed.

### 5. Region V, hyperbolic permittivity region with closed and open IFC polaritons

This region (Fig. 3g) corresponds to Re($\epsilon_x$) > 0, $-1 <$ Re($\epsilon_y$) $< 0$ (hyperbolic dispersion with the $\epsilon_y$ component being near-zero) and $|$Re($\epsilon_x$)Re($\epsilon_y$)$| > 1$. In this case, two different sets of physical solutions to the dispersion relation, Eq. (4), emerge.

The first set combines to two different solutions of Eq. (4) into a unique, closed, isofrequency curve. For angles $0 < \alpha < \theta_2$, $\rho^-(\alpha) \in \mathbb{I}^+$ tracing a hyperbola, while the atan($\xi$) term is positive imaginary. For its complementary set of angles $\theta_2 < \alpha < \frac{\pi}{2}$, $\rho^+(\alpha) \in \mathbb{R}^+$ and atan($\xi$) $\in \mathbb{R}^+$. Interestingly, the overall dispersion $q(\omega, \alpha)$ form a *single* and *closed* mode with $l = 1$ (see Supp. Mat. for details) since the branch cuts of the $\rho$ term and the arctangent term cancel out. A similar closed mode has been previously reported in the context of near-field heat transfer [69]; here, we note that this mode corresponds to an ENZ polariton, as confirmed by the negative phase velocity (calculated in the Supp. Mat.) and the associated electric field distribution obtained in this work (Fig. 4a).

The second set corresponds to $\rho^-(\alpha) \in \mathbb{R}^+$ a hyperbola, atan($\xi$) $\in \mathbb{R}^+$ and $l \geq 2$. The IFCs are open with asymptotes corresponding to $\rho^- \to \infty$ at an angle $\theta_1$ given by Eq. (5). These modes correspond to hyperbolic volume polaritons, each showing multiple nodes in $E_z$: $N_{nodes}(E_z) = l + 1$ (similar to Fig. 4c). The momenta of the modes increase with increased mode index and its phase velocity is negative. The HVMs with concave wavefronts propagate along the *x*-axis with an aperture angle given by $\theta_1$, while for all angles a mode propagates (the ENZ mode) indicating the simultaneous propagation of open and closed IFC polaritons.

Hence, this region supports the simultaneous propagation of a family of open-IFC HVMs, and a single closed IFC ENZ mode, both exhibiting negative phase velocity (Fig. 3h).

### 6. Region VI, hyperbolic permittivity region with open IFC polaritons

This region (Fig. 3g) corresponds to Re ($\epsilon_x$) > 0, Re ($\epsilon_y$) < $-1$ (hyperbolic permittivity). Here, we find three different families of modes.

The first one corresponds to $\rho^-(\alpha) \in \mathbb{R}^+$ a hyperbola, atan($\xi$) $\in \mathbb{I}^+ < 1$ and $l = 0$. It corresponds to an open IFC mode with a vertex at $\alpha = \frac{\pi}{2}$, an asymptote angle given by Eq. (6), and positive phase velocity. The field profile of these mode exhibits one node in $E_z(z)$, see Fig. 4b.

The second one corresponds to $\rho^-(\alpha) \in \mathbb{I}^-$ a hyperbola, atan($\xi$) $\in \mathbb{I}^+ > 1$ , and $l = 1$. It corresponds to an open IFC with a vertex at $\alpha = 0$, an asymptote angle also given by Eq. (6) but with negative phase velocity. The field profile of these mode exhibits two nodes in $E_z(z)$.

The third one corresponds to $\rho^-(\alpha) \in \mathbb{R}^+$, atan($\xi$) $\in \mathbb{R}^+$ and $l \geq 2$. These solutions correspond to a family of open IFCs with a vertex at $\alpha = 0$, an asymptote given by Eq. (5) and negative phase velocity. The momenta of these modes increase with increasing mode number $l$ and $N_{nodes}(E_z) = l + 1$. This family of modes is similar to that of HVMs in region V indicating that they don't undergo a topological transition at the $\epsilon_y = -1$.

The electric field over the metasurface (Fig. 3l) shows various modes propagating along the *x*-axis, and a mode of larger wavelength propagating along the *y*-axis. Observing those fields is challenging to conclude that the asymptote direction is different for the modes of $l \geq 2$, a careful analysis is made in the supplementary material.

*javier.alfaro@unavarra.es

†inigo.liberal@unavarra.es

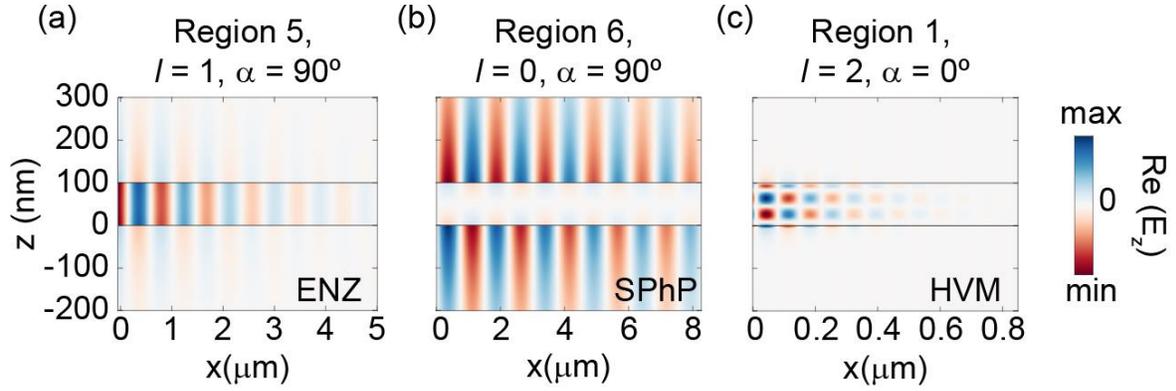

Figure 4. **Electric field distribution of selected polaritonic modes**. Cross sectional distribution of the electric field $\text{Re}(E_z)$ for representative polaritonic modes from different regions, calculated at the points indicated in Fig. 3g. In panels (a) and (b) the propagation direction is $\alpha = 0$, while in panel (c) it is $\alpha = \frac{\pi}{2}$. In all panels, energy propagates from left to right. The phase velocity of the mode is negative in panel (a). The $x$-axis is scaled differently in each panel to enhance visualization.

Summarizing, in this region we find two modes ($l = 0,1$) corresponding to hyperbolic surface polaritons [68,70] that occupy complementary angles with respect of its asymptote $\theta_2$ and have opposite phase velocities. Interestingly, the family of modes $l \geq 2$ have a different asymptote, given by $\theta_1$. Thus, in this region we have three distinct families of modes, all with open IFCs (one with vertex at $\alpha = \frac{\pi}{2}$ and the other two at $\alpha = 0$), having different asymptote angles.

### 7. Region VII, negative permittivity with anisotropic surface phonon polaritons

This region (Fig. 3j) corresponds to $\text{Re}(\epsilon_x) < 0$, $\text{Re}(\epsilon_y) < -1$ and $|\text{Re}(\epsilon_x)\text{Re}(\epsilon_y)| > 1$ (all-negative $\hat{\epsilon}$). In this region we find only one mode, corresponding to $\rho^+(\alpha) \in \mathbb{I}^+$ an ellipse, $\text{atan}(\xi) \in \mathbb{I}^- < 1$ and $l = 0$. The number of nodes $E_z(z)$ is one (similar to that of Fig. 4b), and the cross-sectional field profile corresponds to that of the SPhP. The phase velocity of the mode is positive, and the field induced by a point dipole over the metasurface (Fig. 3m) demonstrates polariton propagation in all directions. This region hence supports uniaxial SPhPs.

### IV. DISCUSSION

Negative permittivity arises in Regions II, IV, and VII, corresponding to regions of small filling factors. Regions II and VII support uniaxial ENZ and SPhP respectively while Region IV uniquely accommodates both types of modes at complementary angles, exhibiting open IFCs while preserving their distinct field distributions and phase velocity directions. By comparison, Region I cover all filling factors near the LO phonon, implying that even small perturbations to the SiC layer can readily suppress the ENZ polariton and support HVMs, a unique characteristic emerging from a vanishing permittivity.

As detailed earlier, ENZ modes appear in regions II, IV, and V; however, they exhibit open IFCs only in region IV in which $\hat{\epsilon}$ is all-negative. In that region the IFC opening obeys a different mechanism than the usual in hyperbolic regions: the branch cut of the geometric factor $\text{atan}(\xi)$ of Eq. (4) at $\xi = i$. Conversely, the same mechanism allows the ENZ mode to form a closed IFC in the hyperbolic-dispersion Region V.

In the hyperbolic $\hat{\epsilon}$ regions (I, V, and VI), each region's sign configuration dictates the polaritonic behaviour. Region I, where $\text{sign}(\text{Re}(\hat{\epsilon})) = (-, +, +)$, supports only HVMs with open IFCs, vertices at $\alpha = 0$ and $v_p > 0$, and no ENZ polaritons. Whereas Regions V and VI, with $\text{sign}(\text{Re}(\hat{\epsilon})) = (+, -, -)$, exhibit both open and closed IFCs polaritons, features previously observed in uniaxial hyperbolic systems [68,69], here we explicitly show that these branches present differing phase velocities and field distributions, resulting in a more intricate polaritonic landscape.

*javier.alfaro@unavarra.es

†inigo.liberal@unavarra.es

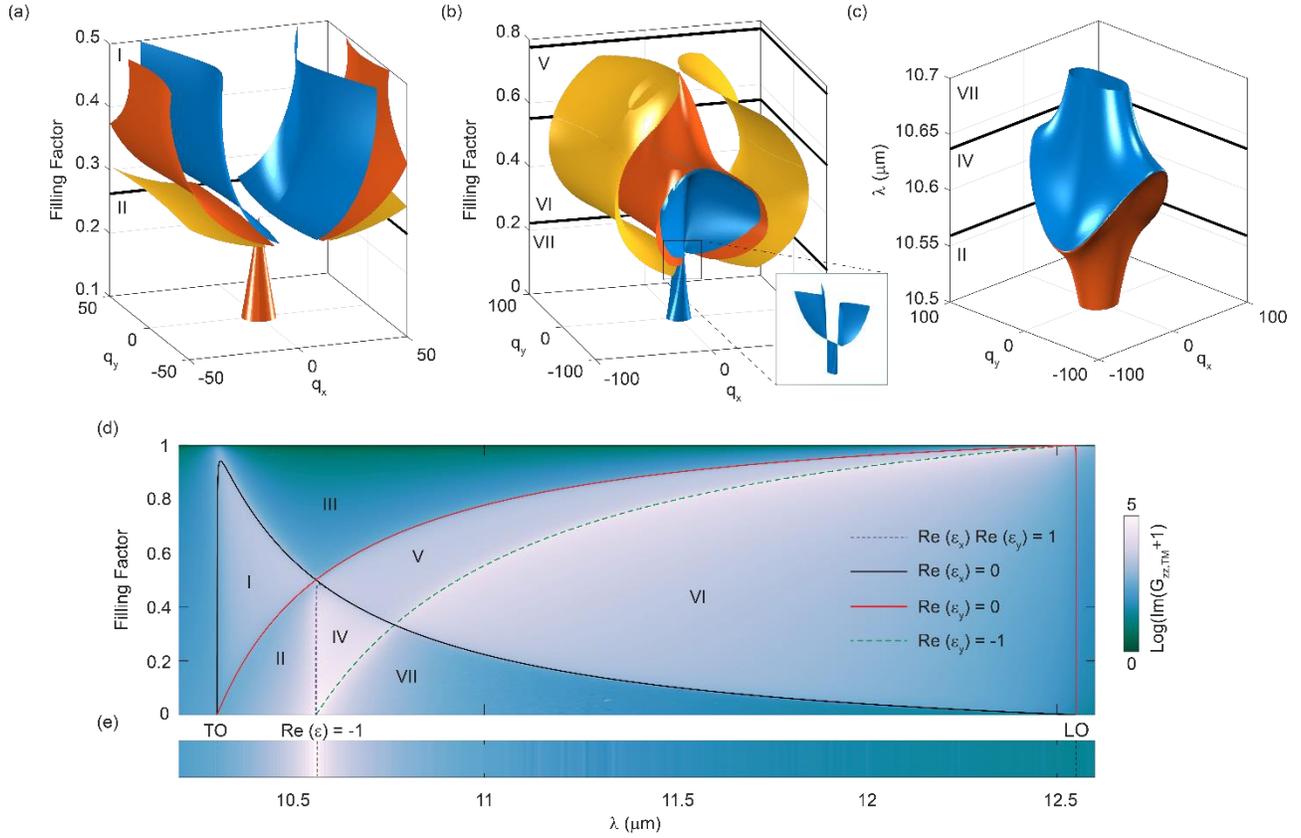

Figure 5. **Polaritonic topological transitions and Purcell enhancement in uniaxial metasurfaces.** (a)–(c) Isofrequency surfaces of the polaritonic modes in three representative regions of the phase diagram. Surfaces are plotted in momentum space $(q_x, q_y)$ as a function of the (a,b) filling factor or (c) wavelength. Blue, orange, and yellow correspond to the $l = 0, 1, 2$ mode indices, respectively. The inset in (b) shows the detailed transition of the $l = 0$ mode at the VI to VII boundary. The losses have been reduced for better visibility of the transitions. (d) Purcell enhancement computed for a vertical dipole placed 100 nm above the metasurface, shown as a function of wavelength and filling factor. Whiter corresponds to larger LDOS enhancement. Lines demarcate the boundaries between the polaritonic regions (e) Purcell enhancement for an isotropic SiC film. Vertical dashed lines indicate the LO and TO phonon frequencies and the condition $Re(\epsilon_{SiC}) = -1$.

As anticipated, the complete polaritonic phase diagram cannot be reproduced by simpler models, such as that of an infinitesimally-thin-layer [46,62]. In this limit, the dispersion relation describing the lowest order momentum modes is:

$$q(\omega, \alpha) = \frac{i}{\alpha_x \sin^2 \alpha + \alpha_y \cos^2 \alpha}, \quad (7)$$

with $\alpha_j = \frac{k_0 t \epsilon_j}{2i}$, the normalized conductivity in the $j$−direction with $j = x, y$. While this model correctly captures the dispersion of the $l = 0$ polariton in most of regions VI and VII (see Supp. Mat.), it fails to reproduce the dispersion of other modes −most notably, the ENZ polaritons. These findings confirm that a 3D approach is essential to capture the complete polaritonic landscape.

At the boundaries between the regions (Table II), topological transitions [51] –qualitative changes in the isofrequency curves– occur. A clear example is the hyperbolic-to-elliptic permittivity transition between regions I and II (Fig. 5a), where the HVM modes in region II collapse to $q_x = 0$, and the ENZ polariton in region I collapses to $q = 0$.

*javier.alfaro@unavarra.es  
†inigo.liberal@unavarra.es

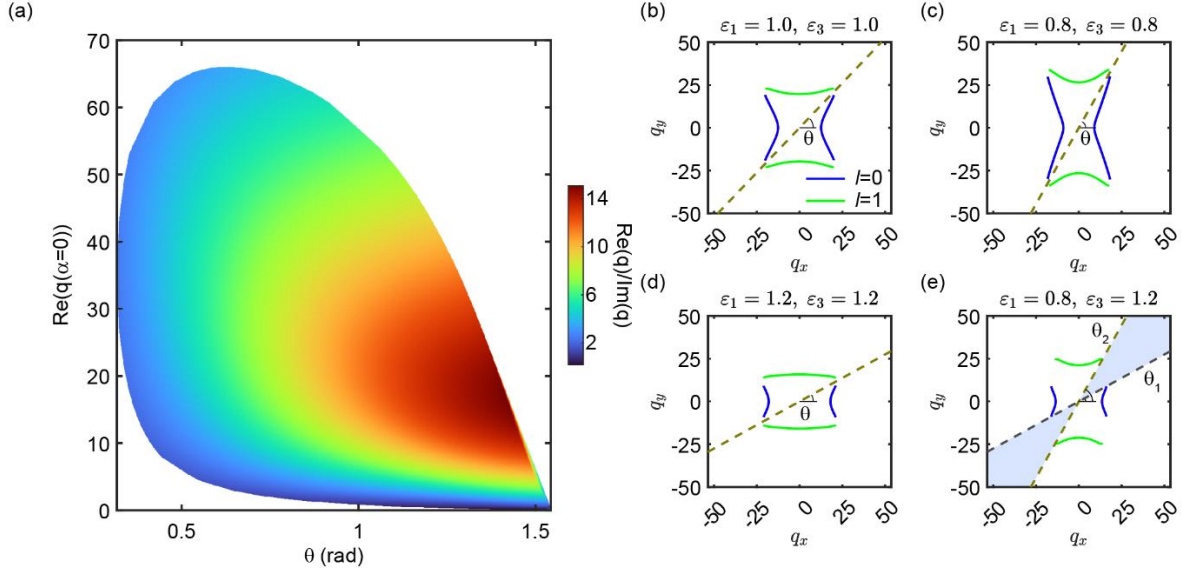

Figure 6: **Tuning of the metasurface**. (a) Pairs of $(\theta, q(\alpha = 0))$ for the $l = 0$ mode across all Region IV, with the color scale indicating the normalized propagation length, $\frac{\text{Re}(q)}{\text{Im}(q)}$. (b-e) asymptote angle variation by adjusting the superstrate ($\epsilon_1$) and substrate ($\epsilon_3$) permittivities, as indicated in each panel title. Panels (b-d) correspond to homogeneous environments where the asymptote for the $l = 0$ and $l = 1$ modes coincide. Panel (e) corresponds to an inhomogeneous environment, where two modes have distinct asymptotes creating a range of angles for which no mode can propagate.

More complex transitions appear at other regions boundaries. As we move from region V to region VI, (Fig. 5b) the $l = 1$ ENZ mode transitions from closed to open IFC. Simultaneously, the $l = 0$ emerges at complementary angles and, as mentioned in Section III.6, higher-order modes remain unaffected. At the VI ↔ VII, the $l = 0$ mode's IFC collapses to a pair of diabolical points [71] transforming subsequently into the closed-IFC SPhP of Region VII. Higher-order modes present in region VI similarly collapse and vanish in region VII.

Topological transitions also occur between all-negative-$\hat{\epsilon}$ regions. At the II → IV boundary, the previously closed IFC of the $l = 1$ mode opens, while the $l = 0$ mode emerges at complementary angles (Fig. 5c). Approaching region VII, the inverse process occurs. The $l = 1$ asymptote angle narrows until it vanishes, which is naturally accompanied by the closure of the IFC of the $l = 0$ mode. In particular, this sequence—from a single closed-IFC mode, to two open-IFC complementary modes, and back to one closed-IFC mode— represents a photonic topological transition not explicitly addressed previously that underscores the fluid interplay between ENZ polaritons and SPhP in anisotropic materials.

The dramatic reshape of the modal landscape at these transitions directly influences the light-matter interaction near the metasurface. To visualize this impact, we numerically calculate the Purcell factor [72,73] for a *z*-oriented electric point dipole 100 nm above the metasurface (Fig. 5d). The Purcell factor exhibits enormous variations at the boundaries between regions, demonstrating the strong influence of the photonic topological transitions on the metasurface properties. As expected, regions with open IFCs modes exhibit broadband Purcell enhancements [74,75], while the Purcell effect clearly diminishes where no polaritonic modes are sustained (region III). We find that the highest Purcell factor occurs along the $\epsilon_x = -1$ curve, being particularly pronounced for the isotropic case (Fig. 5e). This observation challenges the common assumption that hyperbolic permittivity systems yield higher Purcell factors than all-negative $\hat{\epsilon}$. Our calculations illustrate the role of topological mode transitions in the light-matter interaction enhancement of emitters placed nearby.

Controlling the key modal characteristics – specifically, the asymptote angle $\alpha$, and the vertex momentum $\text{Re}(q_0)$– is essential for engineering the polaritonic response of the metasurface. We illustrate the metasurface tunability by varying the filling factor and wavelength and calculating the asymptote angle $\theta$ and the vertex momentum $\text{Re}(q_0)$ for the $l = 0$ mode in Region IV, Fig. 5a. We show that the pair $(\theta, \text{Re}(q_0))$ can be simultaneously tuned over a broad range by changing the filling factor and the wavelength. The colormap indicates the normalized propagation length $\text{Re}(q_0)/\text{Im}(q_0)$. Hence, we demonstrate that by controlling the metasurface design one can access a wide range of asymptote angles and vertex momenta while maintaining relatively low damping, offering a route toward low-loss subwavelength-scale light manipulation.

*javier.alfaro@unavarra.es

†inigo.liberal@unavarra.es

Beyond the intrinsic tunability from the metasurface design, the dielectric environment –the presence of a substrate or superstrate– plays a decisive role in the polaritonic dispersion [76–78], particularly the asymptote angle of the modes. Our previous analysis assumed a symmetric vacuum environment. We find that the asymptote angle defined by Eq. (5) remains unaffected by changes in the surrounding permittivity. In contrast, for modes whose asymptote is described by Eq. (6), the surrounding environment exerts a marked influence: in a homogeneous environment of permittivity $\epsilon_s$, the asymptote angle expression generalizes to:

$$\theta_2 = \arctan\sqrt{\frac{\epsilon_x \epsilon_y - \epsilon_s^2}{\epsilon_s^2 - \epsilon_y^2}}, \quad (8)$$

whereas in an asymmetric setting with permittivities $\epsilon_{1,3}$ it becomes:

$$\theta_2^{+(-)} = \arctan\sqrt{\frac{\epsilon_x \epsilon_y - \max(\min)(\epsilon_1^2, \epsilon_3^2)}{\max(\min)(\epsilon_1^2, \epsilon_3^2) - \epsilon_y^2}}. \quad (9)$$

for the modes in complementary angles (Fig. 6b-e). Here, $\theta_2^+$ corresponds to the $l = 0$ branch and $\theta_2^-$ to the $l = 1$ mode, indicating that the breaking of the up-down symmetry induces an angular bandgap in the polaritonic modes. These expressions remain valid as long as the variations in the surrounding permittivity are sufficiently small to avoid triggering a topological transition in the IFCs [76,79]. Our results show that carefully chosen substrate and superstrate layers provide direct control over the polaritonic asymptote angles, and consequently, the hyperbolic-ray direction [80,81]. Thus, we identify the mechanism by which topological transitions can be driven through dielectric engineering.

The 3D ribbon system, simulated with full-wave numerical methods [65] (see Supp. Mat.), provides important validation of our effective medium theory by confirming the existence of the predicted polaritonic modes across regions I–VII. These simulations successfully reproduce the modes' presence, isofrequency curve topology (open/closed), and qualitatively confirm their phase velocity behavior. However, simulating the full 3D system also reveals significant discrepancies from the homogenized model, highlighting that each ribbon configuration requires independent fine-tuning to realize the desired polaritonic response. This 3D validation thus confirms our theoretical results and suggests a practical route for their experimental realization.

Finally, we highlight that the analytical methodology developed here can be directly applied to related anisotropic photonic systems, including biaxial crystals [62], heterogeneous layered structures [82,83] and shear-type biaxial metasurfaces [84]. Given that these systems' governing equations closely parallel Eq. (4), their study within our analytical framework is both natural and effective.

## V. CONCLUSIONS

This work fundamentally advances the understanding of polaritonic phenomena emerging at the intersection of anisotropic and epsilon-near-zero photonics. We demonstrate a complex interplay between the material's anisotropic permittivity tensor, whose components exhibit significant variations in sign and magnitude, traversing critical points including ENZ transitions, and the thin film geometry that enables the confinement of distinct polaritonic modes (ENZ, HVMs and SPhPs). This interaction fundamentally reshapes the nature and topology of the supported polaritons. Specifically, by mapping the complete polaritonic phase diagram we reveal a breakdown in the conventional correspondence between isofrequency curve topology and the signs of the permittivity tensor components. This is exemplified by anisotropic ENZ polaritons exhibiting unexpected topologies like open IFCs (in non-hyperbolic regimes) and closed IFCs (within hyperbolic regimes).

In doing so, these findings expand the field of ENZ photonics into different directions. The generalized ENZ polaritons inherit the strong confinement and coupling of ENZ modes, with applications for opto-electronic devices [30], near-field heat transfer [64,65], nonreciprocal and tailored thermal emission [25,26,85]. Similarly, the strong nonlinear response of ENZ modes [4–6] could be harnessed for generalized nonlinear effects. The fast tunability of SiC [86,87] could expand the field of spatiotemporal metamaterials towards MIR frequencies, and towards spatiotemporal near-field effects with open IFC modes. Furthermore, the tunability of the asymptote angle and polariton momenta of the ENZ modes offer enhanced control of light routing at subwavelength scales [49]. Finally, the analytical methodology developed here can be extended to more exotic systems, such as twisted metasurfaces, enabling the exploration of new wave phenomena for enhanced light–matter interactions. We believe these results provide a solid foundation for future research into advanced photonic devices that leverage the unique anisotropic response of ENZ thin films.

*javier.alfaro@unavarra.es

†inigo.liberal@unavarra.es

## ACKNOWLEDGMENTS

The authors thank Prof. Alexey Nikitin and Dr. Kirill Voronin for the useful discussions. This work was supported by ERC Starting Grant 948504 NZINATECH. I.L.

## CONFLICT OF INTEREST

The authors declare no conflict of interest.

*javier.alfaro@unavarra.es

†inigo.liberal@unavarra.es

*javier.alfaro@unavarra.es

†inigo.liberal@unavarra.es

*javier.alfaro@unavarra.es

†inigo.liberal@unavarra.es

*javier.alfaro@unavarra.es

†inigo.liberal@unavarra.es